\documentclass[journal=ancac3,manuscript=article]{achemso}

\usepackage[version=3]{mhchem} 
\usepackage{lineno}
\usepackage{graphicx}
\usepackage{lineno}
\usepackage{setspace}
\usepackage{url} 
\usepackage{textcomp}
\usepackage{ragged2e}
\usepackage[utf8]{inputenc}
\usepackage{amssymb}
\usepackage{xspace}
\xspaceaddexceptions{),.;:!?]}

\newcommand{\figscale}{.8}

\newcommand{\wn}{~cm\textsuperscript{-1}\xspace}
\newcommand{\snom}{s-SNOM\xspace}

\author{Gergely Németh}
  \affiliation[SOLEIL]
  {SOLEIL Synchrotron, L'Orme des Merisiers, RD128, Saint Aubin, France 91190}
  \alsoaffiliation[Wigner]
  {Wigner Research Centre for Physics, 29-33 Konkoly-Thege str. Budapest, Hungary, 1121}
  \email{gergely.nemeth@synchrotron-soleil.fr}

\author{Ferenc Borondics}
  \affiliation[SOLEIL]
  {SOLEIL Synchrotron, L'Orme des Merisiers, RD128, Saint Aubin, France 91190}
  \email{ferenc.borondics@synchrotron-soleil.fr}

\title{Step-scan interferometry for high-fidelity hyperspectral nanoscopy}

\keywords{AFM, SNOM, nano-FTIR, step-scan, infrared, Spectroscopic methods, FTIR, interferometry, machine learning}

\begin{document}

\justifying
\begin{abstract}

 Fourier transform infrared nanospectroscopy (nano-FTIR) is a novel, increasingly adopted characterization method that leverages decades of established knowledge in infrared spectroscopy at the nanoscale. It opens up new possibilities in the characterization of composite materials and nanophotonic systems. Besides the rapid adoption and new possibilities, the nanoscale nature of these measurements poses new challenges for infrared spectroscopy. The current implementations of hyperspectral image acquisition at high spatial resolution suffer from significant artifacts due to thermal instabilities, which heavily affect positioning. As a result, the spatial and spectral fidelity of the measurements can be unreliable for long acquisitions. Here, we propose a new nano-FTIR measurement methodology based on step-scan interferometry and image registration. We demonstrate that the method provides superior spatial fidelity for photonics research and enables the collection of larger datasets, paving the way for bringing machine learning to characterize nanoscale heterogeneity.

\end{abstract}

\section{Introduction}

Research and innovation are closely tied to the maturity of available technologies and methods. For example, scanning probe microscopy (SPM) opened up a new avenue for nanoscale physics. First, scanning tunneling microscopy (STM) was implemented, allowing the spatially encoded recording of tunneling current between a conductive probe and a sample. Later, atomic force microscopy (AFM), where the probe reports mechanical or topography information appeared. \cite{Binnig1986-nl} AFM is an excellent platform for hybrid, correlative experiments that combine the mechanical information directly available from the probe with various other phenomena, such as electrical potential (KPFM) \cite{Nonnenmacher1991-cv}, magnetism (MFM) \cite{Martin1987-hz, Saenz1987-ni} or light (NSOM). \cite{Novotny2007-ar}

As a result of decades of development, today's AFMs have very high performance. However, even in advanced instruments, thermal stability, heat management, vibrations or the hysteresis of piezoelectric components pose a significant challenge and their combined effects can lead to sample drift relative to probe position. Experiments that require slow scanning, i.e. long integration of the signal at each observed position, can suffer from such effects and sometimes might even become impossible. Continued efforts by manufacturers and scientists focus on both instrumental solutions, such as mechanical stability and drift correction during scanning, and on improving interpretability by correcting recorded data after the measurement.\cite{Degenhardt2022}

One of the more recent optical SPM modalities, the subject of this work, is scattering-type near-field scanning optical microscopy (\snom), which is a powerful technique to investigate photonic phenomena and material properties at the nanoscale. The idea was proposed as early as 1928 \cite{Synge1928-ux} with various realizations of measurement setups spreading from the visible to the THz range. \cite{Hillenbrand2025-ib} Commercially, \snom setups became available in the late 2000s following the groundbreaking work of Hillenbrand and Keilmann. \cite{Knoll2000, Ocelic2006, Huth2012} Another AFM-coupled IR technique is photothermal AFM-IR, invented by A.~Dazzi and his colleagues, where the IR signal is detected through the photothermal expansion of the sample upon IR excitation. \cite{Dazzi2005}

Infrared-coupled AFM experiments, both \snom and photothermal AFM, come in two flavors. In imaging mode, they record the spatial distribution of IR excitation at a given wavelength, and spectroscopic characterization can be performed by tuning the excitation wavelength through repeated imaging. The collected multispectral image stack can be corrected for drift using the mechanical channels (e.g., deflection signal, tapping amplitude, tapping phase, etc.) simultaneously recorded by the AFM. However, when using broadband sources, such as broadband lasers or synchrotron facilities, the signal is recorded using a Michelson interferometer in a single point or in a raster map.~\cite{Nemeth2024, Bechtel2014, Huth2012, Amenabar2017, Donaldson16} In such cases, the tip must remain in a fixed position relative to the sample while recording the interferogram, thus sample drift can significantly influence the measurements and limit the size of objects that can be reliably measured as well as the area, spatial resolution and fidelity of hyperspectral maps.

\begin{figure}[htbp]
\centering\includegraphics[width=.8\linewidth]{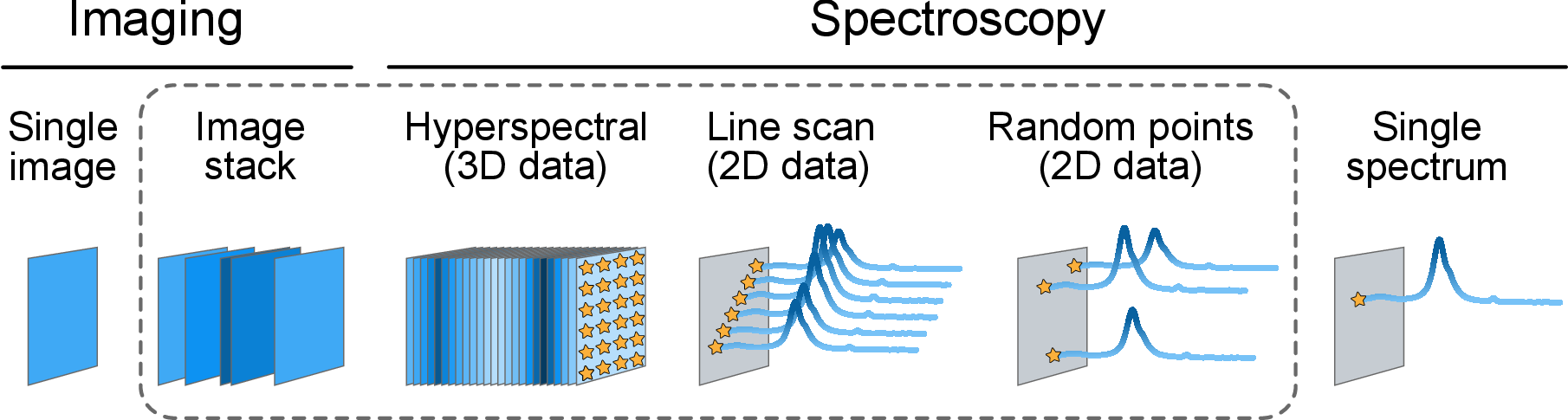}
\caption{Measurement modalities in hyperspectral data collection. Color intensity and gradients represent the intensity of an absorption feature.}
\label{fig:modalities}
\end{figure}

A sample drift-distorted measurement could range from a single-point experiment that requires long-term stability of the probe (e.g., measuring a single nanoparticle) to the acquisition of a high-resolution hyperspectral dataset.\footnote{In this work we will refer to AFM-based infrared spectroscopy measurements, but the principles and the developed methods are generalizable for any AFM experiment where a physical parameter is scanned slowly (e.g. temperature, gating voltage, etc.).} On Fig.~\ref{fig:modalities}, we highlight the typical spectroscopy data collection modalities ranging from a single image acquisition to a spectrum collection at a single point. The sample can move during spectrum acquisition in random directions, effectively leading to measurements of different parts of the sample within the same spectrum acquisition. This would practically invalidate the data, as even a before-and-after measurement to verify position stability doesn't allow for drift correction, as the drift can be random and nonlinear. Therefore, drift is a serious problem for any sample with nanoscale heterogeneity, when the stage drift exceeds the lateral size of the structure, or when the distorted image could lead to misinterpretation of the data (see Methods Section).

Here, we describe a solution for the first time, to our best knowledge, and solve the problem of sample drift in \snom/nano-FTIR measurements. Our method, based on the step-scan technique of data collection, allows for pixel-perfect datasets through aligning the hyperspectral data cube using the mechanical signal channels recorded simultaneously with the optical signal in each interferometer position.

\section{Results and Discussion}
\subsection{Step-scan nano-FTIR}

\begin{figure}[htbp!]
\centering\includegraphics[width=.8\linewidth]{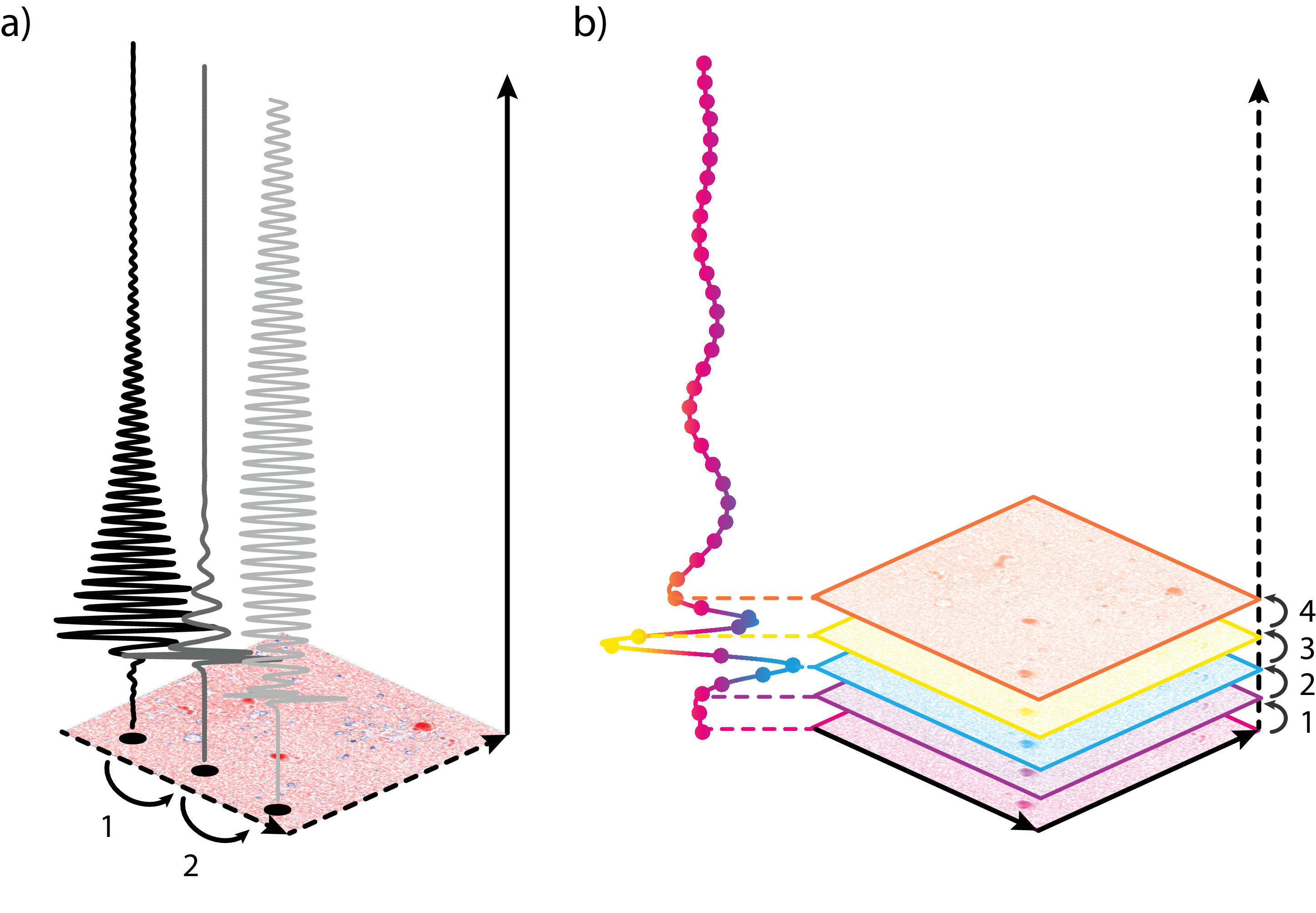}
\caption{Schematics of measurement methodology for rapid-scan (a) and step-scan (b) hyperspectral imaging. Dashed arrows represent the fast scanning directions, while the solid arrows show the slow axis. As shown in (a), a new interferogram is measured at each spatial point. In step-scan modality b), a new image is measured at each position along the optical path of the scanning mirror.}
\label{fig:datacube_record}
\end{figure}

The method we propose here aims to minimize the effect of long-term instabilities on the final dataset by swapping the data acquisition of the measurement directions. Instead of sweeping the moving mirror of the interferometer and trying to keep the sample in the same position, we apply the step-scan interferometric measurement approach. \cite{Manning1993} We fix the scanning mirror position and capture a conventional \snom image. For the next image, we move the interferometer mirror to the next position and repeat the image acquisition. The mirror moves step-by-step after each image until it reaches the optical path difference (OPD) that provides the required spectral resolution. At the end, the recorded 2D images can be assembled into an interferometric data cube with the optical retardation as the third dimension, which will provide a hyperspectral data cube after Fourier transformation to obtain spectra at each point of the scanned area. Fig.~\ref{fig:datacube_record} shows a schematic depiction of the difference between traditional continuous-scan nano-FTIR and our step-scan nano-FTIR approach. We note that throughout our work, we will refer to continuous-scan as point-by-point \emph{rapid-scan} method.

At the advent of FTIR instrumentation step-scan was used to record interferograms, which later transformed into applications for time-resolved FTIR and FTIR imaging using focal-plane arrays (FPA). Step-scan FTIR imaging with FPAs and step-scan nano-FTIR using \snom instruments are similar concepts except that for nano-FTIR the data is not recorded instantaneously for all the pixels. Rather, \snom images are captured by raster scanning. The timespan of a hyperspectral step-scan nano-FTIR measurement is equivalent to the a point-by-point rapid-scan hyperspectral measurement with the same number of pixels in the data cube. With the step-scan method, the relative drift between the sample stage and the AFM probe persists, but it manifests as an offset or distortion between consecutive images.

\begin{figure}[htbp!]
\centering\includegraphics[width=\linewidth]{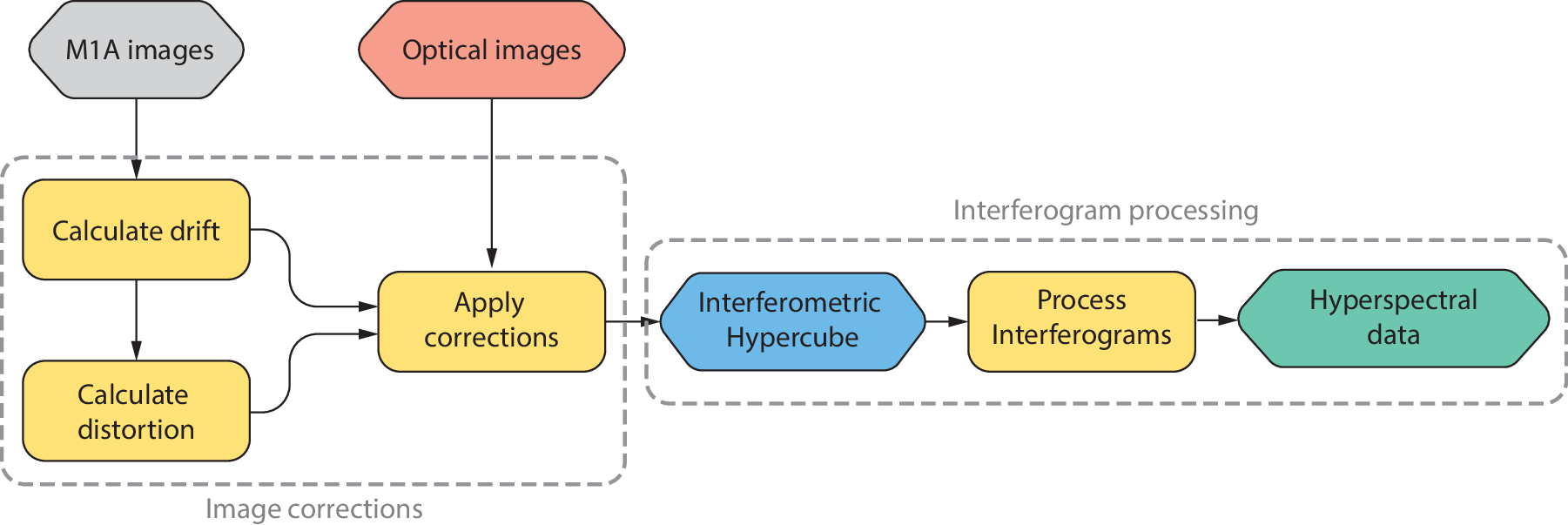}
\caption{Flowchart of the processing approach. M1A images are used to compute image transformations, which are then applied to the simultaneously acquired optical images. The corrected images are organized into a data cube representing the interferograms. By processing the interferograms, we obtain a Hyperspectral data cube.}
\label{fig:flowchart}
\end{figure}

However, this drift or distortion can be corrected by image registration algorithms with a post-processing workflow shown in the flowchart in Fig.~\ref{fig:flowchart}. An advantage of the step-scan nano-FTIR over conventional step-scan interferometry is that we acquire not only optical images but also AFM topography. By the simultaneous acquisition, the topography and optics match pixel-perfect. Since the optical contrast in an image can change drastically with the interferometer position, it is cumbersome, and in some cases impossible, to use optical images for the alignment procedure. However, the mechanical does not change during the measurement and is therefore perfect for tracking drift between images. In practice, we found that the M1A (tapping amplitude image) data yield the least amount of artifacts in the registration results. As we show in Fig.~\ref{fig:flowchart} we use the M1A images to calculate image correction including drift and distortion. As the next step, we apply the corrections to the optical images and lint them up into an interferometric hypercube. In the last step, we process the asymmetric interferograms as usual to get the final hyperspectral data. \cite{Nemeth2024}

\subsection{Experimental realization}
\label{sec:experimental}

To demonstrate the principles described above and subsequently apply them on real world research problems, we used a commercially available near-field microscope (NeaSCOPE IR+, Attocube GmBH, Haar, Germany) equipped with a nanoFTIR module. The commercially available measurement software supports only point-by-point rapid-scan hyperspectral imaging. To implement step-scan data collection, we used the manufacturer's software development kit to independently control each component of the nano-FTIR module and the AFM. The microscope was automated to record a new image at the predefined interferometer mirror positions. Since the AFM is operating in forward-backward raster scanning mode, we collect two images per measurement, by averaging the forward and reverse images, we can further increase the final signal quality.

\begin{figure}[htbp]
\centering\includegraphics[width=.8\linewidth]{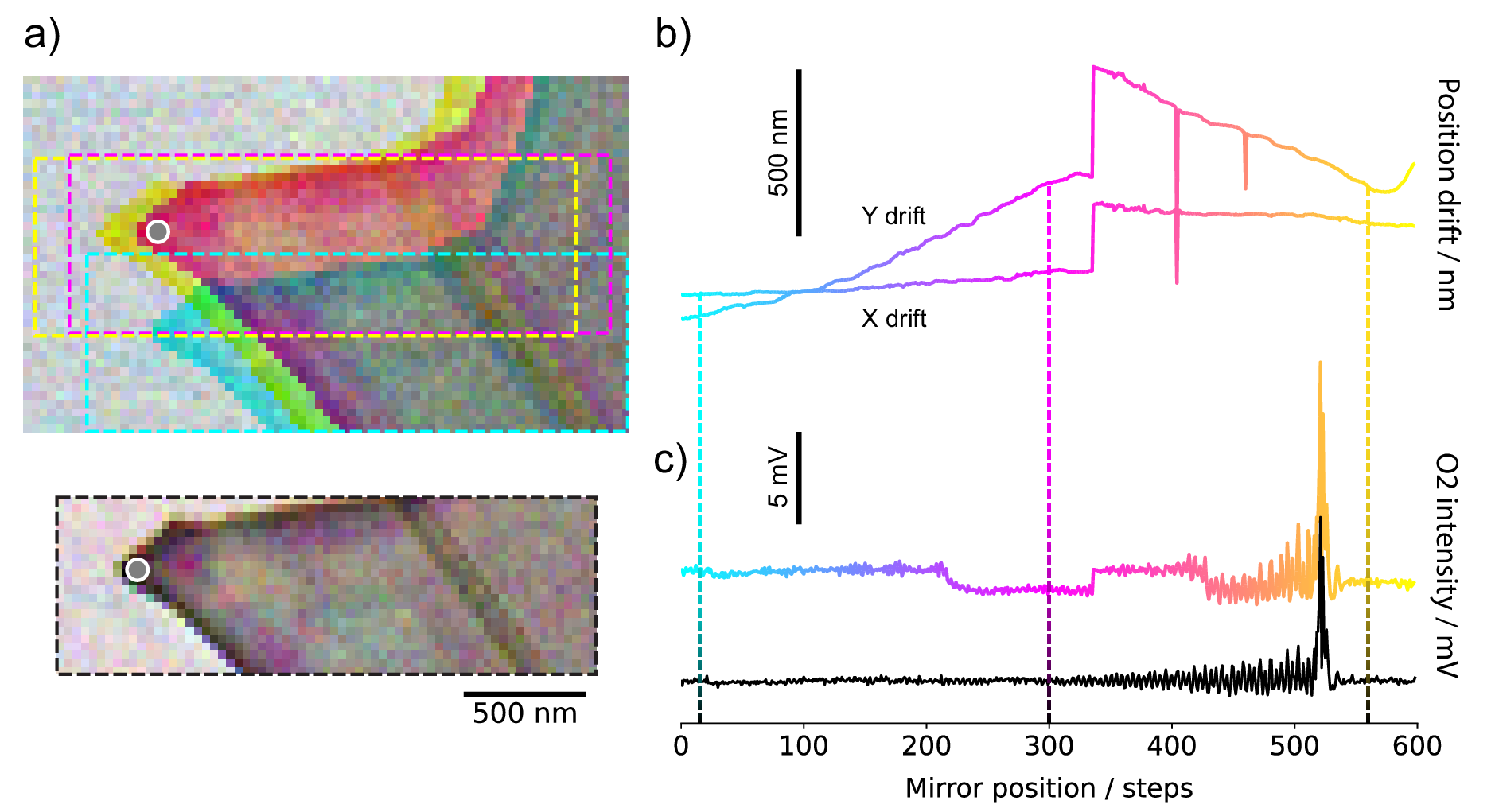}
\caption{Data alignment procedure. a) composite CMY image of three selected frames from the raw dataset showing the drift. Colored dashed frames mark the calculated maximum overlap areas and their physical locations in the images. b) composite image of the three aligned and cropped frames. c) shows, in CMY color gradient, X and Y direction drifts of the raw images and an extracted interferogram from the gray point in a) and b); the black interferogram is taken at the same location from b). Colors are consistent throughout the figure to relate the positions of the frames in the drift and interferogram curve.}
\label{fig:driftcorrection}
\end{figure}

In the following, we present a typical dataset obtained by a step-scan measurement. During the measurement, we recorded a step-scan dataset of a hexagonal boron-nitride (hBN) flake that is deposited on an undoped silicon substrate. Fig.~\ref{fig:driftcorrection} a) shows three slices from the dataset of 600 images as a composite CMY image before and after image drift correction. In the top image, before correction, the composite image shows the evident drift of the hBN flake relative to the captured area. The maximum overlap area between all 600 images is plotted as a frame following the drifting crystal with the same color as the respective image slice. The lower panel in a) shows the same three images after realignment and cropping to the size of the overlapping area. In this example the reduction of the data size after cropping is significant, which depends on the AFM stability during the measurement, and thus, on the magnitude of the sample drift. However, we note that it can be minimized and corrected by calculating the drift on-the-fly during image acquisition, which is another advantage of the step-scan method. Here, we did not apply live tracking on purpose to showcase the effect of drift on the captured data.

Fig.~\ref{fig:driftcorrection} b) presents the sample movement in the X-Y directions visualized by the drift curves, and reaching almost 1~µm. Dashed vertical lines mark the positions of the selected frames in a) with the corresponding color. In~Fig.~\ref{fig:driftcorrection} c) we show an interferogram extracted from the spot marked by the gray circle in a) before (colored curve) and after (black curve) aligning the dataset. Also evidenced by the interferogram curve with the color gradient the tip is drifting on and off of the hBN flake and the interferogram becomes a mixture of responses from the substrate and the hBN making it impossible to interpret the final results. After drift correction,  the extracted interferogram correctly captures the response of the hBN flake at the highlighted position, without any artifacts. In the following, we further compare the advantages and disadvantages of step-scan and rapid-scan nano-FTIR spectroscopy, as it is important to distinguish their applications and domain of validity.

\subsection{Step-scan vs rapid-scan nano-interferometry}

Here, we demonstrate the spatio-spectral stability of step-scan nano-FTIR interferometry and compare it to the conventional rapid-scan approach. We study both the spatial and spectral stability in both cases.

\subsubsection{Spatial stability}\label{sec:spatial_stability}

To showcase and better understand the effect of long-term sample drifts on the final hyperspectral image, we choose to measure a 2.5$\times$1.5~µm hyperspectral map of an hBN flake with different thicknesses and shapes. We acquire the same amount of data in the same timeframe for both methods. The image size was 75$\times$45 pixels, and we sampled the interferogram at 600 points, with a total OPD of 490~µm, yielding $\approx$10\wn spectral resolution. Since the step-scan method always produces forward-reverse images, we also acquired two interferograms at each point for the rapid-scan tests to match the data-acquisition times of the two methods. After each measurement of around 14 hours, we applied the data processing pipeline described in the previous sections as appropriate to get the final hyperspectral data cubes. 

\begin{figure}[ht]
\centering\includegraphics[width=.8\linewidth]{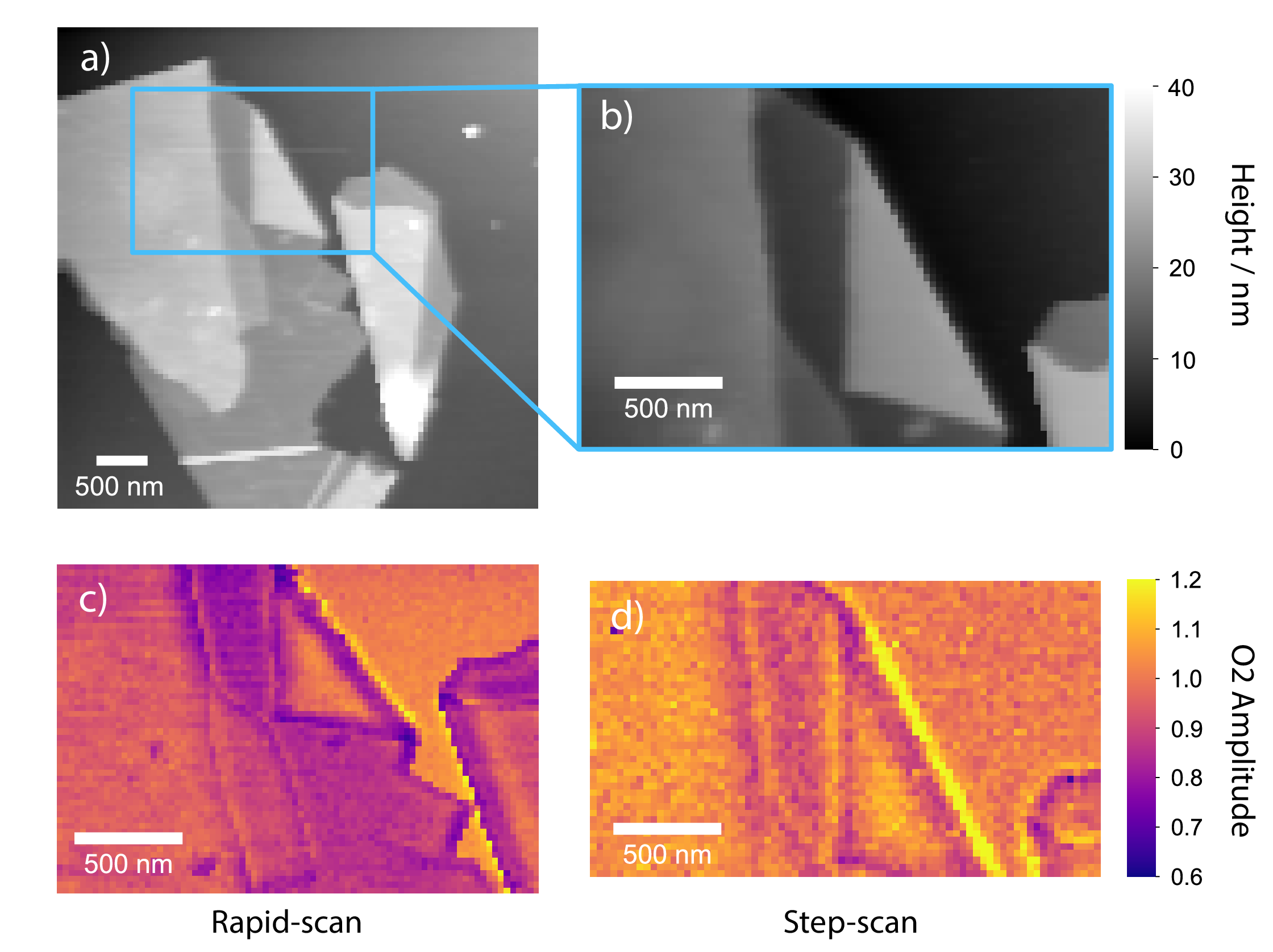}
\caption{Effect of stage drift on hyperspectral imaging. a) an AFM topography overview of the hBN flakes chosen for hyperspectral imaging. The blue rectangle shows the specific area (plotted in b) that was selected up to measure a c) point-by-point rapid-scan and d) step-scan hyperspectral dataset. Both c) and d) show a spectral image slice at 800\wn. We note that d) is slightly smaller than the original area due to cropping to the overlapping region of the drifted images.}
\label{fig:fig-hBN-distortion}
\end{figure}

In Fig.~\ref{fig:fig-hBN-distortion} we compare the spatial accuracy of step-scan versus rapid-scan acquisitions. a) shows an AFM overview of the sample with the blue rectangle marking the selected area for subsequent hyperspectral measurements. A ground-truth AFM image of the selection is presented in b) for better inspection. After the measurement and the necessary processing steps (drift correction and FFT for step-scan, FFT for rapid-scan) c) and d) show a spectral image slice at 800\wn (near the out-of-plane phonon mode of hBN) from both hyperspectral datasets. The images reveal a striking difference between the results from the two methods. The step-scan image provides AFM image accuracy with a perfect match. In contrast, the rapid-scan slice yields a severely distorted image, with recognizable but highly altered spatial features of the hBN sample. The main triangular fold of the flake is non-uniformly squeezed, and other parts of the sample shift into the field of view. This distorted image is clearly a consequence of a sample drift of roughly 1~µm.

The introduced distortion has serious consequences in the interpretation of the spectral results. For example, in 2D materials, where surface polariton interferometry is a common technique in \snom to study polaritonic excitations, the wavelength of the polariton interferences at the edges can be completely misinterpreted, leading to false results. The same issue is present in simple analytical studies of nanoscale heterogeneities or multiphase materials. The size of the different domains can be misread from the rapid-scan hyperspectral results.

Only the step-scan method introduced above provides truthful spatial information, an essential requirement for real-space, high-fidelity nanoscale measurements.

\subsubsection{Spectral stability}

So far, we have focused primarily on eliminating consequences of long-term instabilities in the spatial variables of spectral imaging. However, spectral stability and consistency is equally important. Assigning the slow acquisition to the spectral axis we put the requirements of long-term stability to the interferometer.

It was already shown in conventional step-scan interferometry that the position error of the OPD is crucial to achieve high signal-to-noise ratio. \cite{griffith_book_snr} Specifically, the effect of position errors on SNR is given as 
\begin{equation}
    SNR = 4/(\Delta\delta\cdot\sigma)
\end{equation}
where $\Delta\delta$ is the error in OPD and $\sigma$ is the spatial frequency of the radiation in wavenumbers. \cite{griffith_book_snr} This formula works for the average OPD error, assuming the same average error at each mirror position. Another issue, similar to the spatial position of the AFM tip earlier, is that the drift of the interferogram mirror could vary over time. As a result, the interferogram might be locally phase-modulated when the reference mirror drifts. This depends mostly on the construction of the interferometer, but major manufacturers have invested decades of engineering in improving the thermal stability of their interferometers. The other advantage of step-scan nano-FTIR is that, even if we assume a similar reference mirror position drift as with the sample stage, the 0.5-1~µm drift would occur over 14 hours along the full scan distance of 800~µm (in our experiments). This drift has less serious consequences, although it is important regarding the spectral signal-to-noise.

\begin{figure}[ht]
\centering\includegraphics[width=\linewidth]{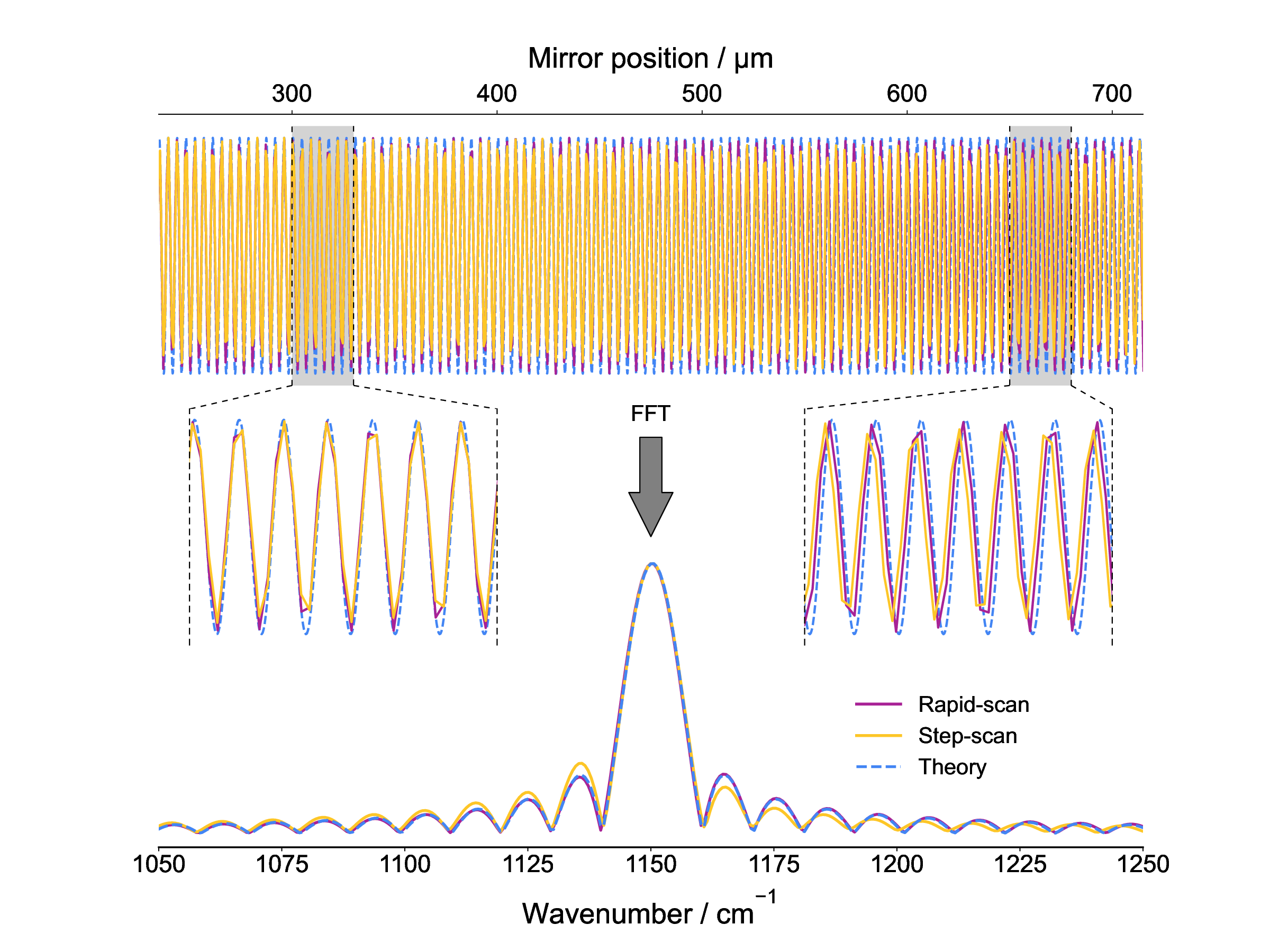}
\caption{Stability characterization: Single-wavelength interferograms measured for 1150\wn. Rapid-scan (purple) and step-scan (yellow) data are overlaid on the theoretical sinusoid. The insets help to visualize the overlap, and the FFT spectra prove that both techniques provide results close to theoretical interferograms with some negligible modulation.}
\label{fig:fig-qcl-stability}
\end{figure}

To test the long-term stability of our system and compare spectral performance, we conducted step-scan and rapid-scan measurements using a quantum cascade laser tuned to 1150\wn. As a result, the ideal interferograms in both methods have to be perfect sinusoids representing the given frequency. Any instabilities in the position control of the scanning mirror result in phase modulation of the perfect sinusoid. In Fig.~\ref{fig:fig-qcl-stability}, we show the similarity of rapid-scan, step-scan interferograms with overlaying the theoretical expectation. The two insets show zoom-ins of the interferograms near the end (I) and at the beginning (II) of the measurements. The interferograms overlap almost entirely except for a small part at the beginning of the scans. To better understand the effect, we also plot the Fourier spectrum of the three signals. The spectra show a slight asymmetry in the side lobes of the step-scan data. The interferograms were captured in the same way as all of our measurements described in the Methods Section.

\begin{figure}[htbp]
\centering\includegraphics[width=\linewidth]{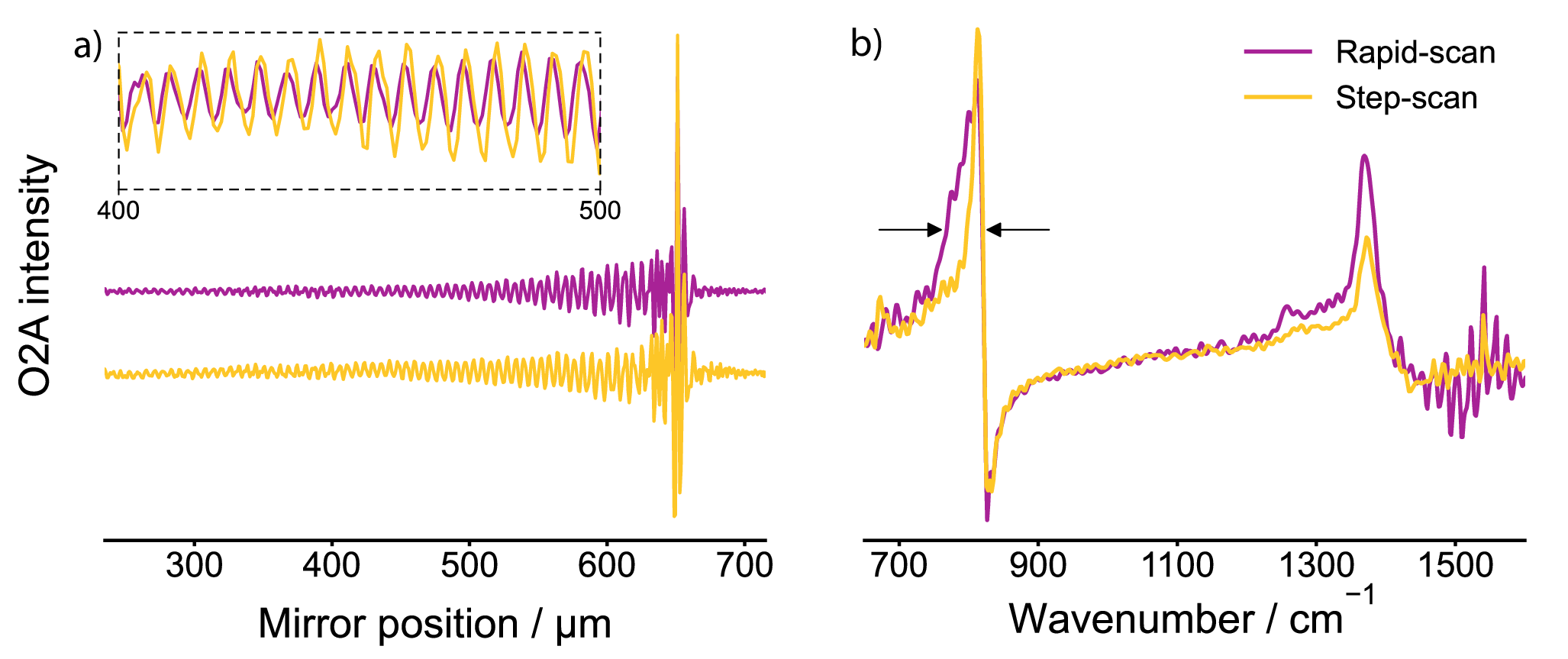}
\caption{Comparison of step-scan and rapid-scan data for the second-harmonic optical amplitude signal (O2A). a) Raw rapid-scan and step-scan interferogram offset for clarity. The inset shows agreement between the two methods at long optical path differences. b) Corresponding spectra obtained after the Fourier transform of a). Both spectra were normalized by the average spectrum of the silicon substrate in the same measurement.}
\label{fig:fig-hBN-stability}
\end{figure}

To further benchmark the two methods, we compared the interferograms and spectra from the measurements shown in Fig.~\ref{fig:fig-hBN-distortion}. We plot the second harmonic interferograms and spectra in Fig.~\ref{fig:fig-hBN-stability} a) and b), respectively. The inset in a) shows a zoom in on the marked section of the interferograms. The extended oscillations arise from the low-frequency polariton band of the hBN around 800\wn. This is an intense phonon line with low damping, thus the long signature in the interferograms. Both spectra were extracted from the spot marked in Fig \ref{fig:fig-hBN-distortion}. The spectra from the two methods show slight differences. Marked by the arrows, the rapid-scan spectrum presents a broadening of the low-frequency hBN phonon. Since we tested our system multiple times and found that both methods yield similar spectral performance in our setup, we attribute this to sample drift during the rapid-scan. As a consequence, the location picked from the two measurements is not exactly the same.

Based on these tests, we can conclude that the step-scan method provides a matching spectral performance to that of the rapid-scan method. However, we note that it has an increased sensitivity to the OPD relations between the two arms of the spectrometer.

\subsubsection{Other sources of spectral noise}

In our study, we primarily focused on the stability of the mirror position control during long acquisitions, but other factors also affect the OPD and spectral noise. One of these factors is the changes in the atmosphere of the interferometer. As water absorption at ambient conditions is usually a problem in infrared spectroscopy, most instruments are purged with dry gases or kept in vacuum. The stability of the purge gas affects the final spectra in two ways: If the purge is not homogeneous between the two arms of the interferometer, it results in a refractive index difference, which will change the OPD relations. The other potential problem with purging, and more generally with the atmosphere, arises from changes in water and CO$_2$ concentrations over time, resulting in differences in the absorption lines of the reference and sample measurements.

We note that both rapid- and step-scan methods are prone to these problems due to their long acquisition times, and that this needs to be addressed in the same way in both cases. A summary of the noise sources can be found in Refs.~\citenum{Manning1997,griffith_book_snr}.

\subsection{Applications}

With the development of step-scan nano-FTIR we aim to make nanoscale hyperspectral imaging as accurate as possible. It can be beneficial for applications that require high spatial fidelity or high acquisition accuracy. Here, we present examples of scientific studies and demonstrate that the step-scan nano-FTIR is the only method suitable for obtaining results that were previously impossible.

\subsubsection{Polariton interferometry}

Polariton interferometry is a unique capability of \snom, enabling the visualization and study of hybrid states of light and matter called polaritons. It has been used in a tremendous number of studies that unravel the existence and properties of nanoscale polaritonic phenomena. \cite{Basov2016,Basov2020} The key concept that makes \snom so successful for investigating polaritons is its ability to visualize the standing-wave pattern of surface waves reflected from material boundaries and inhomogeneities. The interference pattern allows the extraction of the polariton wavelength at the excitation energy used in the experiment. In the traditional approach, researchers acquire \snom images of their nanoscale system (e.g., a flake of a 2D material) at different energies and extract a wave-pattern profile. This is tedious work and requires a lot of manual labor. Also, it is limited by the emission range of the excitation laser.

Hyperspectral measurements with broadband sources are more reliable and convenient for studying surface waves. Broadband lasers and especially synchrotrons provide light in an ultra-broad energy range. \cite{Bechtel2014} In recent studies, point-by-point hyperspectral linescans are applied to measure the spectra along a line defined perpendicular to the material boundary. From the final dataset, the profile for each energy can be extracted. \cite{hbn1,Dolado2022,Ma2018} However, the accuracy of the method depends on how the line across the edge of the flake was placed and, as we have seen above, on the amount of sample drift during the measurement.

For structures with complex geometries, full hyperspectral maps are needed to unveil the polariton interference patterns and localized mode fields. As seen in Fig.~\ref{fig:fig-hBN-distortion}, the point-by-point rapid-scan approach is not capable to accurately capture the spatial relations of the sample. As a consequence, the extracted field profiles can be inaccurate, leading to misinterpretation and incorrect conclusions. In contrast, step-scan  nano-FTIR measurements can easily provide the required spatial accuracy and spectral quality.

The highest spatial accuracy is required to measure polariton interference in quasi-one-dimensional objects. Metallic single-walled carbon nanotubes (CNTs), for example, can host strong plasmon excitation in the mid-infrared range. \cite{Shi2015} Luttinger-plasmons in CNTs show extreme wavelength and mode confinement. Literature data only show \snom measurements conducted at a handful of wavelengths using tunable lasers in a limited frequency range, thus the broadband frequency response of these plasmons is still unknown. Nano-FTIR measurements could provide a better insight into the nature of 1D plasmons, but due to their small size of $\approx$\,1~nm, it is impossible to measure a line profile along the nanotube axis using the conventional point-by-point rapid-scan method because of sample drift. The AFM would need to have better than 1~nm stability during the entire measurement that can last for hours.

\begin{figure}[htbp]
\centering\includegraphics[width=\figscale\linewidth]{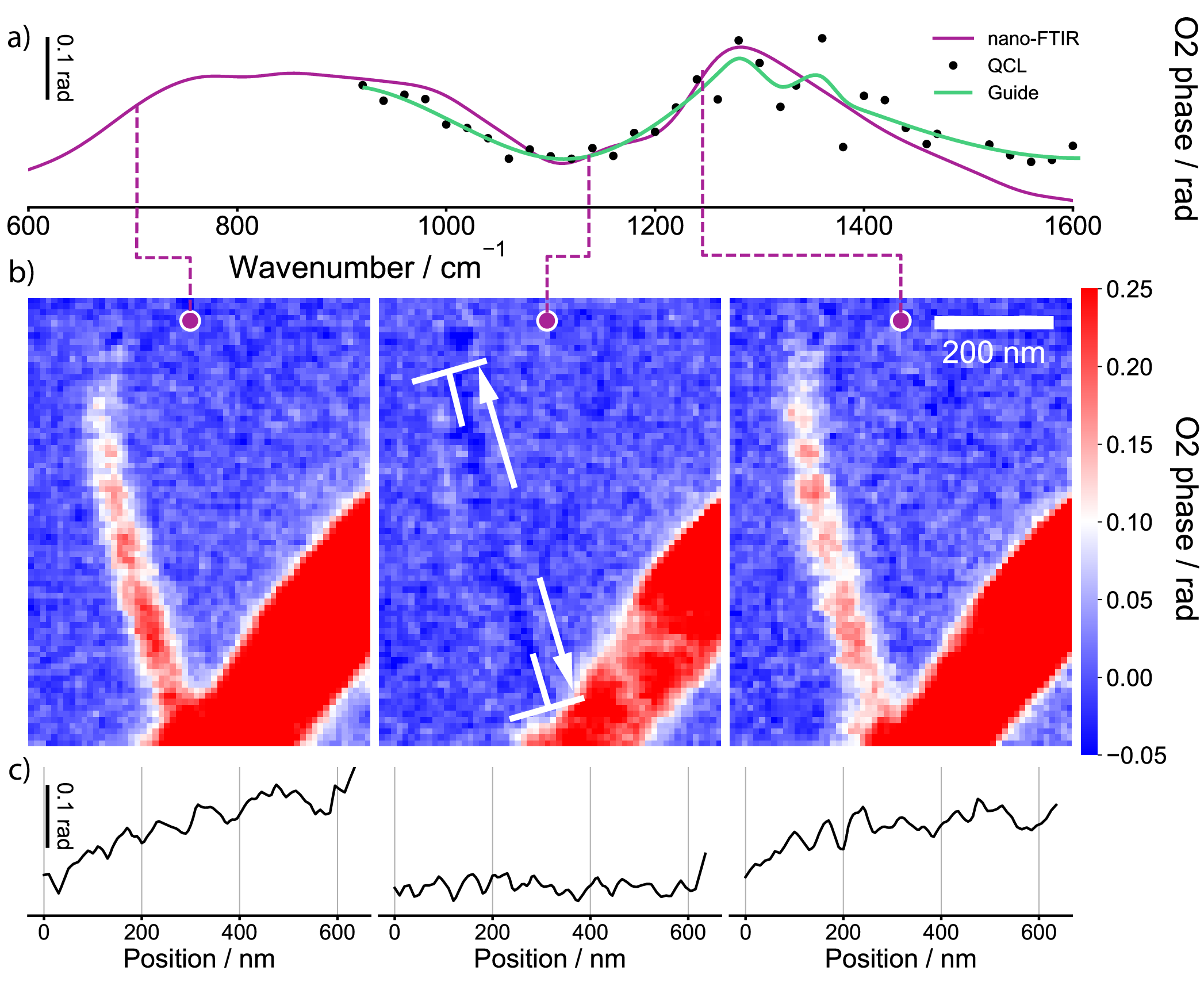}
\caption{Hyperspectral step-scan polariton interferometry on an individual carbon nanotube. a) The second-harmonic phase spectrum of the nanotube shown below. b) Phase images for the three different wavelengths indicated by the dashed purple lines. c) Line profiles extracted along the nanotube axis marked by the white arrows in the middle image of b)}
\label{fig:cnt}
\end{figure}

Here, we show that step-scan nano-FTIR is capable of capturing such accurate details and provides unprecedented information about the spectral characteristics of carbon nanotube plasmons. Fig.~\ref{fig:cnt} presents our results obtained by measuring an individual carbon nanotube. We conducted the step-scan nano-FTIR mapping on a small piece of an individual nanotube with a diameter of 1.4~nm. The imaged area, which is also shown in Fig.~\ref{fig:cnt}, is 600~nm~$\times$~800~nm with a pixel size of 10~nm. The sample drift during the measurements was much larger than the selected area, so it could easily shift too much to be corrected in post-processing. To solve the issue, we automated our system to track the nanotube by recalculating the drift after each image acquisition and starting the next image from the drift-compensated coordinates. This way, we could maximize the overlap area of the final image stack.

Fig~\ref{fig:cnt} a) shows the average phase spectrum derived from the phase values of the nanotube shown in the images in b). The spectrum shows the typical spectra of nanotubes on Si. It presents a broad plasmon response with a wide dip at around 1100\wn, which originates from the ultrastrong coupling between the nanotube plasmons and the phonons of the native silica layer on the substrate. \cite{Nemeth2022} The images show single-wavelength slices extracted at the spectral positions marked by dashed purple lines. The standing wave patterns arising from the interference of the tip-launched and the reflected plasmons are clearly visible both in the images and the line profiles along the nanotube plotted in c).

Previous studies could capture only a portion of the spectral characteristics of the nanotube plasmon band due to limitations in laser imaging. \cite{Nemeth2022,FengWangResonators,Tian2018} With the step-scan method, we could record the full bandwidth of mid-infrared 1D plasmonic excitations in carbon nanotubes from 600\wn, which is unprecedented to date, and highlights the ultrabroad nature of nanotube plasmons. As the images show, they maintain a long propagation length across the entire excitation range, rendering nanotube plasmons extremely attractive candidates for enhanced infrared detection at the nanoscale. Additionally, these results demonstrate the potential of step-scan nano-FTIR technique and its ability to facilitate novel research.

For further examples of polariton interferometry, we invite the reader to watch the supplementary animation, which shows a step-scan hyperspectral polariton interference measurement of hexagonal boron nitride (hBN). It demonstrates the quality and fidelity of hyperspectral data captured by step-scan nano-FTIR.

\subsubsection{Machine learning for biospectroscopy}

Machine learning or AI tools are extremely useful and can provide unique insight in general and in the natural sciences as well. However, statistical methods require large amounts of data to make sensible analysis, and multivariate statistics and machine learning algorithms are even more data-hungry for reasons of model training and making meaningful predictions. Such tools are commonly used in biospectroscopy, where data is plentiful, and it is also often required to collect large, statistically relevant datasets. Now, with the use of the step-scan data collection we introduce here, such datasets are available to uncover hidden details at the nanoscale using various machine learning methods.

The step-scan nano-FTIR measurements were done on \textit{Lingulaulax polyedra}, a modern thecate dinoflagellate cell. \cite{Mertens2023, Bellefeuille2014, Head2024} Thin sections of the cells were deposited on a gold substrate. A 5~µm~$\times$~4.5~µm area with 60~nm step size was set up for the hyperspectral measurements with the same interferometer parameters described in the Methods Section.

\begin{figure}[htbp]
\centering\includegraphics[width=\linewidth]{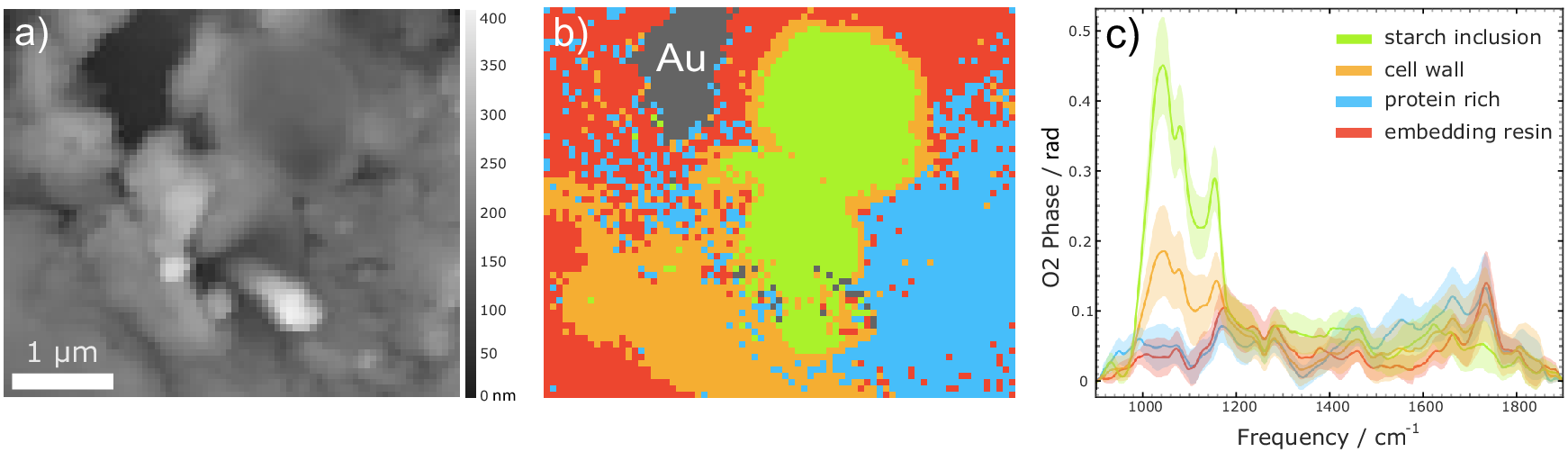}
\caption{Hyperspectral data processing on a biological sample thin section. a) AFM topography image, scale in nanometers; b) hyperspectral clustering of t-SNE projection scores; c) cluster average spectra with corresponding colors to b). Note that spectra in the gray cluster labeled Au were removed from the dataset to avoid distortion in the data analysis.}
\label{fig:fig-pjotr-clusters}
\end{figure}

Here, we briefly introduce a workflow aimed to discover regions in the sample built with Quasar \cite{Toplak-Borondics2017, Toplak-Borondics2021}, an open-source machine learning software built for the natural sciences, on a thin TEM section of a single-cell organism. The sample has a tear in the thin section that exposes the gold substrate, which can be used as a spectral reference area. After a complex FFT and frame alignment, we isolated the region corresponding to the substrate using k-means clustering and used the average spectrum for spectral normalization. Subsequently, we projected the dataset using the t-SNE \cite{t-SNE-Hinton} method and clustered the resulting scores with hierarchical clustering (normalized Euclidean distances, average linkage, four final clusters). Fig.~\ref{fig:fig-pjotr-clusters}~a) and b) show the topography and the resulting cluster distributions, respectively.

This processing workflow reveals four distinct regions, where the main spectral differences show the accumulation of starch (green cluster, high intensity between 1100 -- 1200\wn) and the cell wall (orange cluster, lower intensity between 1100 -- 1200\wn especially compared to the $\approx$1175\wn peak), a resin permeated, potentially protein rich region (blue cluster, characteristic amide bands at $\approx$1550, and 1650\wn), and the embedding resin (red cluster, strong spectral feature $\approx$1750\wn). 

Previous studies were only using single-wavelength \snom images or a few-point nano-FTIR spectroscopy to study cell composition and organelles. \cite{Marxer2026, Giliberti1016, Greaves2023, deCarvalho2024, Kanevche2021} In a recent study, F.~Marxer et \textit{al.} addressed interferometer drifts to achieve high-spectral-quality nano-FTIR measurements. \cite{Marxer2026} We note that our method is less prone to this problem (as shown in the Spectral stability Section) because the reference and sample points are measured simultaneously at the same mirror positions. To the best of our knowledge, the above-presented hyperspectral dataset and its analysis workflow are the first of their kind in the literature and were made possible with the new step-scan technique introduced in this paper, clearing the way for real subcellular hyperspectral nanoscopy.

The Quasar workflow and the corresponding dataset are available in our dedicated GitHub repository. \cite{step-scan-repo}

\subsubsection{Further applications}

We note that the step-scan technique is transferable to other nanoscale spectroscopy involving an interferometer. The same principle can be applied to broadband photothermal AFM microscopy \cite{Donaldson16}, or nanoscale THz time-domain spectroscopy. \cite{Aghamiri2019} It also opens up the possibility to conduct sample-modulated time-resolved \snom/nano-FTIR measurements, bringing a decades-old technique to the nanoscale. Furthermore, it can also extend the possibilities of novel nanoscale-pump-probe measurements. \cite{Esses2026} Additionally, our correction workflow can be applied to multispectral \snom narrow band laser imaging or other AFM modalities where one parameter is stepped and consecutive AFM images are collected. The alignment procedure and image distortion correction make it possible to correct multispectral \snom images for negative phase artifact without the need for expensive multi-wavelength measurement setups. \cite{Niehues2023,Vicentini2023}

\section{Conclusion}

We have introduced a new method to perform nano-FTIR measurements and collect large-area hyperspectral datasets with nanometer-level accuracy based on the combination of step-scan interferometry and image registration. The core idea is to assign the fast collection to the spatial scanning and acquire the spectral dependency on the slower timescale. We showed that using topography images as alignment frames for image registration enables correction of optical data and results in pixel-perfect alignment, yielding unprecedented data quality and reliability at every spatial location. We demonstrated the power of the new method by applying it to real scientific problems. We presented a broadband spectroscopic study on carbon nanotube Luttinger plasmons from 600\wn, which is not possible with other techniques. We also showed, for the first time, the use of advanced machine learning tools on hyperspectral nano-FTIR datasets. The concepts and data analysis methods described here open new possibilities for nanoscale spectroscopy and are generalizable to other SPM modalities while providing distortion-free, high spatial-fidelity data.

\section{Methods and Experimental Procedures}\label{sec:methods}
\subsection{nano-FTIR measurements}

We used a commercially available near-field microscope (NeaSCOPE IR+, Attocube GmBH, Haar, Germany) equipped with a nanoFTIR module where the reference mirror of the interferometer can move up to 800~µm distance. The interferometer uses a piezo-controlled closed-loop positioner to drive the reference mirror. The commercially available measurement software supports only point-by-point rapid-scan hyperspectral imaging. To implement step-scan data collection, we used the manufacturer's software development kit to independently control each component of the nano-FTIR module and the AFM.

For all of our measurements presented in this paper (if not stated otherwise), we used an interferometer scanning distance of 490~µm, which results in 980~µm of maximum retardation, thus $\approx$10\wn spectral resolution. We automated the data acquisition to capture 600 images, a forward-reverse pair at each mirror position. Thus, the mirror steps were set to 0.8167~µm, achieving the maximum frequency of 3056.12\wn. We note that in nano-FTIR, the resolution depends on the location of the white-light position (WLP) relative to the scanned retardation window and the exact apodization function used to process the interferograms. The WLP positioning can vary widely, but we mostly followed our recently published guide in Ref.~\citenum{Nemeth2024}.

\subsection{Carbon nanotubes samples}

The sample was made from high-purity metallic carbon nanotubes purchased in buckypaper form from NanoIntegris (IsoNanotubes-M M95\% Thick Film). After cutting a millimeter-size piece from the film, we sonicated it in 20 mL toluene for 40 min to break the nanotube bundles and create a dilute, homogeneous dispersion. After sonication, we quickly layered 2-3 drops of the suspension on top of distilled water in a vial. The toluene does not mix with water and remains on top as a thin layer. We dipped a clean, undoped silicon chip into the water and slowly pulled it through the suspension layer. The nanotubes adhered to the substrate as small patches containing individual nanotubes and small bundles.

\subsection{Dinoflagellates}

Several cells were embedded in resin and subsequently ultramicrotomed into 100~nm-thick sections. The sections were then deposited on gold substrates. Cell sections with well-preserved cell organelle shapes were first identified from simple AFM images. More details about sample harvesting and preparation can be found in Refs. \citenum{Mertens2023,Bellefeuille2014,Head2024}.

\begin{acknowledgement}
This project was supported by an ANR/DFG binational grant: UltraSNOM (ANR-23-CE42-0030, DFG Project-ID529998081 \textit{“Ultrasensing in the nearfield: polariton enhanced molecular fingerprinting”}) and by the National Research, Development and Innovation Office of Hungary (NKFIH, project No. OTKA K 143153). The authors thank P.~Meyvisch and K.~Mertens for providing a sample to demonstrate the step-scan nano-FTIR technique and subsequent machine learning analysis possibilities on biological materials as well as B.~K{\"a}stner, S. Fairman and D.~Siebenkotten for their collaboration on live image tracking.
\end{acknowledgement}

\begin{suppinfo}

Animation demonstrating sample drift during measurements, animation showcasing a full hyperspectral polariton interferometry dataset on hexagonal boron nitride. 

\end{suppinfo}

\bibliography{ref}

\newpage

\begin{figure}[htbp]
\centering\includegraphics[width=3.25in]{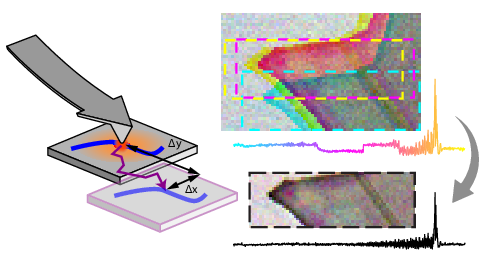}
\caption{For Table of Contents Only}
\label{fig:toc}
\end{figure}

\end{document}